\documentclass[sigconf,nonacm,natbib=false]{acmart}
\makeatletter
\global\@ACM@balancefalse
\makeatother
\usepackage{balance}

\setcopyright{none}
\renewcommand\footnotetextcopyrightpermission[1]{}
\usepackage[
  backend=biber,
  style=numeric,
  sorting=none,
  giveninits=true,
  maxbibnames=3,
  minbibnames=3,
  doi=false,
  isbn=false,
  url=false,
  eprint=false
]{biblatex}
\DeclareNameAlias{author}{family-given}
\DeclareNameAlias{editor}{family-given}

\renewrobustcmd*{\bibinitperiod}{}
\renewrobustcmd*{\bibinitdelim}{}
\DeclareDelimFormat{multinamedelim}{\addcomma\space}
\DeclareDelimFormat{finalnamedelim}{\addcomma\space}
\DefineBibliographyStrings{english}{andothers={et\addabbrvspace al\adddot}}
\DeclareFieldFormat[article,inproceedings,incollection,thesis,misc]{title}{#1}
\DeclareFieldFormat{pages}{#1}

\AtBeginBibliography{\sloppy}

\newbibmacro*{compact:author-title}{%
  \printnames{author}%
  \setunit{\addperiod\space}%
  \printfield{title}}

\DeclareBibliographyDriver{article}{%
  \usebibmacro{bibindex}\usebibmacro{begentry}%
  \usebibmacro{compact:author-title}\newunit\newblock
  \printfield{journaltitle}\setunit{\addcomma\space}\printfield{year}%
  \iffieldundef{volume}{}{\setunit{\addcomma\space}\printfield{volume}%
    \iffieldundef{number}{}{\printtext{\mkbibparens{\printfield{number}}}}}%
  \iffieldundef{pages}{}{\setunit{\addcolon\space}\printfield{pages}}%
  \usebibmacro{finentry}}

\DeclareBibliographyDriver{inproceedings}{%
  \usebibmacro{bibindex}\usebibmacro{begentry}%
  \usebibmacro{compact:author-title}\newunit\newblock
  \printfield{booktitle}\setunit{\addcomma\space}\printfield{year}%
  \iffieldundef{pages}{}{\setunit{\addcolon\space}\printfield{pages}}%
  \usebibmacro{finentry}}

\DeclareBibliographyDriver{thesis}{%
  \usebibmacro{bibindex}\usebibmacro{begentry}%
  \usebibmacro{compact:author-title}\newunit\newblock
  \printtext{Thesis}%
  \iflistundef{institution}{}{\setunit{\addcomma\space}\printlist{institution}}%
  \setunit{\addcomma\space}\printfield{year}\usebibmacro{finentry}}

\DeclareBibliographyDriver{misc}{%
  \usebibmacro{bibindex}\usebibmacro{begentry}%
  \usebibmacro{compact:author-title}%
  \iffieldundef{howpublished}{}{\setunit{\addperiod\space}\printfield{howpublished}}%
  \iffieldundef{year}{}{\setunit{\addcomma\space}\printfield{year}}%
  \usebibmacro{finentry}}

\title{Profiling and Evolution of Intellectual Property}

\author{Bowen Yu}
\affiliation{%
  \institution{School of Computer Science (National Pilot School of Software Engineering), Beijing University of Posts and Telecommunications; Beijing Key Laboratory of Intelligent Telecommunication Software and Multimedia}
  \city{Beijing}
  \country{China}}

\author{Yingxia Shao}
\authornote{Corresponding author: shaoyx@bupt.edu.cn.}
\affiliation{%
  \institution{School of Computer Science (National Pilot School of Software Engineering), Beijing University of Posts and Telecommunications; Beijing Key Laboratory of Intelligent Telecommunication Software and Multimedia}
  \city{Beijing}
  \country{China}}

\author{Ang Li}
\affiliation{%
  \institution{School of Computer Science (National Pilot School of Software Engineering), Beijing University of Posts and Telecommunications; Beijing Key Laboratory of Intelligent Telecommunication Software and Multimedia}
  \city{Beijing}
  \country{China}}

\begin{abstract}
In recent years, with the rapid growth of Internet data, the number and types of scientific and technological resources are also rapidly expanding. However, the increase in the number and category of information data will also increase the cost of information acquisition. For technology-based enterprises or users, in addition to general papers, patents, and other resources, policies related to technology or the development of their industries should also belong to a type of scientific and technological resource. Extracting valuable science and technology policy resources from a huge amount of mixed-content data and providing accurate and fast retrieval will help break down information barriers and reduce information-acquisition costs, which has profound social significance and utility. This article focuses on the difficulties and problems in the field of science and technology policy and introduces related technologies and developments.
\end{abstract}


\begin{document}
\maketitle

\section{Introduction}

In recent years, with the rapid growth of information data on the Internet, the number and types of scientific and technological resources have rapidly expanded~\cite{yang2021ipResourcePortrait}. The increase in the number and category of information data sometimes increases the cost of information acquisition. For any individual's value standard, disorganized data means that a large amount of information is not of interest and that more time is required to identify valid information. Multi-view clustering illustrates how heterogeneous scientific information can be organized through complementary subspaces~\cite{xue2019deepLowRank}. For scholar-oriented resources, dynamic interest tracking can further connect multi-view clustering with the evolution of researchers' interests~\cite{li2023dynamicScholarClustering}. Taking technology-based enterprises as an example, in addition to papers and patents, policies related to science and technology or supporting the development of their industries are also scientific and technological resources. Such resources are mixed with a large amount of irrelevant policy data, increasing acquisition costs and difficulty. Interpretable machine-learning models are valuable in such decision-support settings because they make automated judgments easier for users to understand~\cite{li2019interpretableDecision}. Extracting valuable resources from mixed-content data and providing accurate and fast retrieval can break down information barriers and reduce acquisition costs.

Policy resources usually come from multiple fields and disciplines. The characteristics of multiple data sources lead to inherent difficulties in collecting and obtaining policy resources. Web-content mining surveys summarize the diversity of extraction strategies required for these sources~\cite{shah2022webContentMining,pujar2021webContentTools}. Sequential market-state modeling is another example of extracting structured signals from large, noisy online data collections~\cite{zhao2017marketState}. Sentiment-variation-aware analysis can similarly explain abrupt sentiment spikes in temporally evolving public-event data~\cite{li2025sentimentSpike}. Community-detection methods based on deep modularity optimization offer a complementary way to reveal coherent groups in complex information networks~\cite{yang2016deepModularity}. When learning is distributed across these sources, dynamic client selection and adaptive gradient compression can reduce federated communication costs~\cite{pan2025rfcsc}. For different Internet information sources, different collection rules must be set, and a general strategy is needed to reduce labor costs. In addition to the initial collection of web-page data, the body text is the main content of policy resources. Accurate text extraction is both a means of obtaining policy resources and a prerequisite for subsequent algorithm training. Approaches that preserve textual and non-textual semantics show why page structure and presentation both matter~\cite{zachariasova2014webImaging}.

\section{Web Content Extraction of Science and Technology Policy Resources}

Web pages differ in structure, presentation, and token types, and parameter differences can affect extraction quality. Microsoft Research Asia proposed the vision-based page segmentation (VIPS) algorithm~\cite{cai2003vips}, which combines visual representation with a DOM tree. Geometry-aware visual reconstruction, such as bi-projection fusion for omnidirectional image super-resolution, further illustrates the importance of preserving spatial information in visual resources~\cite{wang2024omnidirectionalSr}. Visual similarity and text-structure analysis have also been studied for phishing-page comparison and automatic summarization~\cite{lv2015visualSimilarity,shen2014webSummarization}. An improved hidden Markov model can be combined with simulated annealing for web information extraction~\cite{li2014simulatedAnnealingHMM}. VIPS-style algorithms perform well on pages with a single visual form and pronounced structural differences, but a page must be completely rendered before analysis, consuming substantial resources.

Another approach is template-based extraction~\cite{barYossef2002templateDetection}. It assumes that pages are constructed with the same or similar templates, so repeated parts are treated as non-body text and differing parts as body content. Template-based extraction and domain-ontology methods extend this idea~\cite{yang2013templateExtraction,gu2014deepWebOntology}, while local density and context modeling provide related anomaly-detection tools~\cite{hu2018anomalyKernel}. In practice, a URL can help determine whether pages share the same structure. However, modeling a separate extractor for each source is labor-intensive, and an extractor must be modified whenever a target website is revised.

Widely used extraction methods design heuristic strategies from HTML information, including text density, synthetic text density, tag ratio, and path ratio~\cite{wu2013pathRatios}. DOM-based content extraction defines text density and combines it with visual importance~\cite{song2015hybridContent}. Entropy-based informative-content density provides another extraction criterion~\cite{annam2016entropyContent}, while paragraph-tag clustering can identify main content in news articles~\cite{carey2016paragraphTags}. Such methods generally score tag nodes with heuristic functions, but a single strategy and manually selected threshold are not highly adaptable across many data sources. Adaptive webpage extraction based on a decision tree instead classifies DOM nodes from extracted features~\cite{lv2019adaptiveDecisionTree}; its effectiveness, however, depends on assumptions about leaf nodes and the set of HTML tags considered.

\section{Text Feature Representation and Classification Methods of Science and Technology Policy Resources}

Computers cannot directly understand text data, so text features must be encoded into a computable form. Feature extraction and representation are important parts of text mining. TF--IDF uses term frequency and inverse document frequency to measure a word's importance in a corpus. It filters common irrelevant words while retaining important ones, but it does not represent word order or position and treats morphological variants separately.

Traditional classification methods require expensive manual feature engineering. Hybrid convolutional-neural-network and nearest-neighbor methods have been explored for short-text classification~\cite{yao2018cnnKnn}, while support-vector-machine feature selection offers a more conventional alternative~\cite{sabbah2018hybridSvm}. For short texts with sparse labels, heterogeneous graph attention can combine document and relation information in a semi-supervised classifier~\cite{hu2019heterogeneousGraphAttention}. Related machine-learning work on constrained state estimation and image translation demonstrates the wider use of learned representations in complex systems~\cite{li2017varianceState,fang2020identityCycleGAN}.

The key to applying deep learning to large-scale classification is to learn text representations and then use structures such as convolutional neural networks (CNNs)~\cite{zhou2020universalityCNN} and recurrent neural networks (RNNs)~\cite{sherstinsky2020rnnLstm}. FastText averages word vectors in a sentence and connects them to a softmax layer, while adding n-gram features to capture local sequence information~\cite{joulin2016fastText,kim2014sentenceCNN}. Its representation remains limited in word-order modeling. CNNs capture local correlations, but a fixed filter field of view can miss longer sequence information. Filter-enhanced MLPs show that frequency-domain filtering can also encode sequential dependencies efficiently without a deep recurrent stack~\cite{zhou2022filterEnhancedMlp}.

RNNs can better express contextual information, and bidirectional RNNs capture variable-length, bidirectional n-gram information. Multi-task recurrent networks have been designed for classification problems~\cite{liu2016rnnMultitask}. TextRCNN combines recurrent and convolutional components~\cite{lai2015textRCNN}; character-level CNNs classify directly from character sequences~\cite{zhang2015charCNN}. Recursive state-estimation research likewise illustrates learning under coupled and uncertain dependencies~\cite{li2017recursiveState}. Hierarchical attention networks assign attention weights to important words and sentences~\cite{yang2016han}. TextGCN performs text classification with a graph containing document and word nodes~\cite{yao2019textGCN}. In a federated heterogeneous-graph setting, reinforcement-based active client selection can prioritize informative participants during representation learning~\cite{wang2025activeClientSelection}. Teacher--student distillation can make graph representation learning more robust when node features or graph structure are incomplete~\cite{huo2023t2gnn}. Heterogeneous-network collaborative filtering further shows how multi-aspect information can be jointly represented~\cite{shi2019deepCollaborativeFiltering}. For scientific publications in particular, semantic-similarity attention combined with hypergraph convolution captures higher-order relations among papers~\cite{li2026semanticHypergraph}.

\section{Text Matching and Retrieval of Science and Technology Policy}

Accurate query processing requires a measure of semantic text similarity and intelligent information retrieval~\cite{yang2015ontologyRetrieval}. Traditional text matching includes bag-of-words~\cite{wang2012bigrams}, TF--IDF~\cite{salton1988termWeighting}, BM25-style probabilistic ranking~\cite{robertson1994twoPoisson}, and Jaccard similarity~\cite{jaccard1912flora}. These techniques mainly solve lexical-level matching and have difficulty with differences in word meaning and structure. Retrieval-oriented masked-autoencoder pretraining provides language representations designed specifically for retrieval tasks~\cite{xiao2022retroMae}.

With deep learning, neural models reduce the cost of feature engineering. Convolutional-pooling latent semantic models improve information retrieval by learning semantic vectors~\cite{shen2014latentSemantic}. Single-semantic models such as DSSM encode two texts independently and calculate their similarity~\cite{huang2013dssm}, while convolutional architectures capture order information in sentence matching~\cite{hu2014sentenceMatching}. Multi-semantic models represent sentences at multiple granularities and consider local features. MV-LSTM generates positional sentence representations with bidirectional LSTMs~\cite{wan2016mvLstm}, and MatchPyramid treats the matching matrix as an image from which a CNN extracts interaction features~\cite{pang2016matchPyramid}. Entity-duet neural ranking further combines entity and document semantics for retrieval~\cite{liu2018entityDuet}. When modalities are distributed across data owners, federated supervised cross-modal retrieval can align representations without directly centralizing the original data~\cite{li2024federatedCrossModal}.

Large text collections also require an efficient indexing and retrieval framework. Boolean queries can be implemented with Lucene~\cite{li2019luceneBoolean}. Because a single machine may not meet the needs of massive corpora, distributed retrieval services are increasingly important. Distributed indexes improve query performance for large-scale data~\cite{dou2014distributedIndex}. Search-engine implementations expose the indexing, matching, and ranking workflow~\cite{vijaya2013webSearchEngine}. Distributed consensus and variance-constrained estimation provide related tools for reliable networked computation~\cite{li2017distributedKalman}. For scientific and technological resources containing several media types, semantics-adversarial and media-adversarial learning can reduce both semantic and modality discrepancies in a shared retrieval space~\cite{li2022semanticsMediaAdversarial}. Federated graph neural networks extend decentralized learning to cross-graph node classification~\cite{guan2021federatedGraph}. Lucene-based Solr and Elasticsearch use distributed indexing, load balancing, failover, and recovery. A distributed intelligent search system based on Elasticsearch can additionally provide recommendation functions~\cite{zeng2016elasticsearchSearch}. FedSIN similarly learns information-network representations through federated self-adaptive learning~\cite{li2026fedSin}. Self-supervised graph co-training provides a related mechanism for learning complementary session representations in recommendation-oriented retrieval~\cite{xia2021graphCoTraining}.

\section{Conclusion}

In view of the characteristics of scientific and technological resources in the policy field under big-data scenarios, this paper summarizes related technologies and progress from three aspects. First, science and technology policy data has inconsistent structures, so extraction methods must support unified processing of multi-source policy data. Second, text-feature representation and classification techniques are needed to model the extracted content. Third, similarity calculation, matching, and retrieval methods are required to deliver relevant policy resources obtained by extraction and mining.

\begin{acks}
This work is supported by the National Key R\&D Program of China (2018YFB1402600) and the National Natural Science Foundation of China (61772083, 61877006, 61802028, and 62002027).
\end{acks}

\balance
\printbibliography[title={References}]
\end{document}